\documentclass{article}

\usepackage{amsmath}
 \usepackage[dblblindworkshop, final]{neurips_2025}
\usepackage[utf8]{inputenc} 
\usepackage[T1]{fontenc}    
\usepackage{hyperref}       
\usepackage{url}            
\usepackage{booktabs}       
\usepackage{amsfonts}       
\usepackage{nicefrac}       
\usepackage{microtype}      
\usepackage{xcolor}         
\usepackage{graphicx,subcaption}
\usepackage[nowatermark]{fixmetodonotes}

\title{Simulation-based inference for neutrino interaction model parameter tuning}
\workshoptitle{Machine Learning and the Physical Sciences (ML4PS)}

\author{%
Karla Tame-Narvaez\\
  Theory Division\\ 
  Fermi National Accelerator Laboratory\\ 
  Batavia, IL 60510 \\
  \texttt{karla@fnal.gov} \\
  \And
  Aleksandra \'Ciprijanovi\'c \\
  Computational Science and AI Directorate\\ 
  Fermi National Accelerator Laboratory\\ 
  Batavia, IL 60510 \\
  Department of Astronomy and Astrophysics\\
  University of Chicago\\ 
  Chicago, IL 60637\\
  \texttt{aleksand@fnal.gov}
  \And
   Steven Gardiner\\
  Computational Science and AI Directorate\\ 
  Fermi National Accelerator Laboratory\\ 
  Batavia, IL 60510 \\
  \texttt{gardiner@fnal.gov}
  \And
   Giuseppe Cerati \\
  Computational Science and AI Directorate\\ 
  Fermi National Accelerator Laboratory\\ 
  Batavia, IL 60510 \\
  \texttt{cerati@fnal.gov}
}

\begin{document}

\maketitle

\begin{abstract}

High-energy physics experiments studying neutrinos rely heavily on simulations
of their interactions with atomic nuclei. Limitations in the theoretical
understanding of these interactions typically necessitate ad hoc tuning of
simulation model parameters to data. Traditional tuning methods for
neutrino experiments have largely relied on simple algorithms for numerical
optimization. While adequate for the modest goals of initial efforts,
the complexity of future neutrino tuning campaigns is expected to increase
substantially, and new approaches will be needed to make progress.
In this paper, we examine the application of simulation-based inference
(SBI) to the neutrino interaction model tuning for the first time. Using
a previous tuning study performed by the MicroBooNE experiment as a test case,
we find that our SBI algorithm can correctly infer the tuned parameter
values when confronted with a mock data set generated according to the
MicroBooNE procedure. This initial proof-of-principle illustrates a promising
new technique for next-generation simulation tuning campaigns for the
neutrino experimental community.
\end{abstract}

\section{Introduction}

A major priority for current research in high-energy physics is the
investigation of the properties of neutrinos. Because these elementary particles
interact weakly with other forms of matter, typical neutrino experiments involve
both intense neutrino sources and large, highly sensitive detectors in order to
record enough reactions for analysis. Beyond the technical demands of the
required apparatus, however, interpretation of the experimental data poses its
own significant difficulties. Among the greatest of these difficulties is the
need for precise simulations of collisions between neutrinos and atomic nuclei.
Despite ongoing efforts within the scientific community, gaining a full
theoretical understanding of these collisions is a formidable problem, and
state-of-the-art simulations of neutrino scattering presently rely on many
rough, semi-empirical approximations.

To obtain simulation predictions that are reliable enough to use for physics
measurements, neutrino experiments have commonly resorted to tuning interaction
model parameters to reference data. The widely-used GENIE simulation
code~\cite{genie2010,genie2021} has been the most popular platform for such
tuning exercises, with several carried out by the developers
themselves~\cite{GENIE:2021zuu, GENIE:2021wox, GENIE:2022qrc, GENIE:2024ufm} as
well as multiple experimental collaborations~\cite{MINERvA:2019kfr,
NOvA:2020rbg, microboonegenietune}. A representative example recently performed
by MicroBooNE~\cite{microboonegenietune} involved using a software framework
called NUISANCE~\cite{nuisance} to tune four GENIE model parameters to neutrino
scattering data previously obtained by the T2K experiment~\cite{T2K:2016jor}. A
satisfactory adjustment to the base GENIE model, dubbed the ``MicroBooNE Tune,''
was ultimately achieved via a simple likelihood fit to the data. However,
pathological results seen during initial attempts led the authors to take the
drastic step of ignoring the reported correlations between bins of the reference
T2K measurement. While this stopgap solution, paired with simple numerical
methods, was adequate for the modest immediate needs of MicroBooNE,
next-generation efforts will be substantially more exacting. To achieve the
stringent precision on neutrino interaction modeling required for the goals of
the field, future tuning campaigns will necessarily involve both larger
parameter spaces and more input data with greater complexity. Technical
innovation in simulation tuning procedures provides a potential solution for the
neutrino community to overcome these looming obstacles.

Although very recently applied to modeling the performance of the detector
hardware in the JUNO neutrino
experiment~\cite{gavrikov2025simulationbasedinferenceprecisionneutrino}, the use of simulation-based inference (SBI) methods for tuning neutrino interaction models has not yet been explored. These methods leverage simulators as a part of the statistical inference procedure, where the goal is to infer the likelihood or posterior distributions for a given experiment or observation. Traditional SBI methods include, for example, Approximate Bayesian Computation~\cite[ABC;][]{Rubin1984} and Approximate Frequentist Computation~\cite[AFC;][]{brehmer2018guide},  which are closely related to the
traditional template histogram and kernel density estimation approaches.
The recent rise of SBI algorithms, which utilize deep
neural networks as surrogates for modeling the conditional probability
densities of the likelihood or posterior distribution in a given inference
problem from simulations, enabled inference even in high-dimensional parameter spaces~\cite{CB2020}. Furthermore, after the training of the deep learning model is performed, the subsequent inference is fast and cheap, i.e., the SBI model is amortized.

Deep learning-based SBI models have already been successfully used in other
areas of physics. For example, in collider physics~\cite{BR2021} for
constraining the Higgs potential for di-Higgs production~\cite{MN2024},
searching for CP violation in leptonic WH production~\cite{BM2024}, measuring
QCD Splittings~\cite{BB2021}, etc. In astrophysics and cosmology, SBI has been
used for the inference of the Hubble constant from binary neutron star
mergers~\cite{GF2021}, the dark matter substructure inference in galaxy-galaxy
strong lenses~\cite{Brehmer_2019, Coogan2022, Anau2023}, inference of strong
lensing parameters~\cite{Legin2021, WagnerCarena2022, Poh2025}, inference of
galaxy properties from spectra~\cite{Khullar2022}, for cosmology inference from
galaxy cluster abundance~\cite{Reza2022}, etc.

In this work, we use SBI~\cite{BoeltsDeistler_sbi_2025} with a neural posterior estimator (NPE) method to revisit the neutrino interaction model tuning
performed by MicroBooNE~\cite{microboonegenietune}. SBI provides an efficient
alternative to conventional methods, as it offers amortized inference, i.e.,
after the upfront cost of training the model is paid, inference can be
performed in seconds. We demonstrate that our SBI model is capable of inferring
correct parameter values when confronted with mock data generated using the
MicroBooNE Tune configuration of GENIE. This technical demonstration reveals
strong potential for near-future use of SBI to obtain parameter values from
actual experimental data sets, thus establishing a novel approach for tuning
neutrino interaction simulations.

\section{Data and Methods}

We simulate neutrino-nucleus collisions using GENIE~\cite{genie2010,genie2021}.
Within the GENIE interaction model, we vary the four model parameters that were
adjusted in the MicroBooNE tune: $\theta_1$ (\texttt{MaCCQE}), $\theta_2$
(\texttt{NormCCMEC}), $\theta_3$ (\texttt{XSecShape\_CCMEC}), and $\theta_4$
(\texttt{RPA\_CCQE}). The former represent the axial mass for CCQE interactions, which governs the $Q^2$ dependence of the cross section. \texttt{NormCCMEC} represents the overall normalization of the CC multi-nucleon (MEC) contribution. \texttt{XSecShape\_CCMEC} represent the shape parameter of the CC MEC cross section. Finally, \texttt{RPA\_CCQE} represents which modifies the Random Phase Approximation correction applied to CCQE processes. More detailed physical meanings of these parameters are described
in detail in the original MicroBooNE publication~\cite{microboonegenietune}.
To provide sufficient coverage of the parameter space in the vicinity of
the best-fit values from the MicroBooNE tune, we created an ensemble of
configurations in which the four parameters were independently sampled
from uniform distributions with the following ranges:
\[
\theta_1 \in [0.961, 1.39]~\text{GeV}, \quad
\theta_2 \in [1.0,\, 3.0] , \quad
\theta_3 \in [0.0,1.0], \quad
\theta_4 \in [0.0, 1.0].
\]

For each configuration of the parameters, we processed the output of GENIE with
NUISANCE~\cite{nuisance} to produce a 58-bin histogram representing a
theoretical prediction that can be directly compared to the ``Analysis I'' T2K
neutrino interaction data set reported in Ref.~\cite{T2K:2016jor}. For the
original MicroBooNE tune, this procedure was used together with a subsequent
likelihood fit to the measured histogram to obtain best-fit parameter values.
In this study, we created a training set of 200,000 configurations and a test
set of 1,000 independent configurations to validate our SBI workflow.

The goal of our algorithm is to infer the underlying set of parameters used
within GENIE from a prediction histogram generated according to the approach
described above. For this task, we utilize the SBI framework developed by
\texttt{mackelab}\footnote{https://github.com/sbi-dev/sbi}, implemented in
\texttt{python}. The model takes as input the four physics parameters,
$\theta_i$, where $i=1,2,3,4$, along with their associated histograms, $x_i$, to learn the inverse
mapping from histograms to parameters.

To facilitate a more efficient learning of the posterior distribution, we use a simple three-layer embedding network to reduce the dimensionality of the histograms from $58$ to $24$ summary features~\footnote{Comparable results can be obtained with lower-dimensional input data; however, we found that the network becomes overconfident when the number of parameters is reduced below ten. Using 24 parameters proved to be a stable choice, maintaining a well-calibrated model. Further studies on architecture design and optimal parameter tuning are ongoing.}. These embeddings are then used as inputs to the NPE. The NPE uses a Masked Autoregressive Flow~\cite{papamakarios2018maskedautoregressiveflowdensity} architecture with six transformations and $55$ hidden features in each block. Both the embedding network and the MAF are trained together to allow the embedding network to learn the most informative summaries, which will lead to the best posterior predictions.

For the training loop, we use mini-batches of size $512$, a learning rate of $10^{-2}$, and a training/validation split of $90\%/10\%$ to monitor and mitigate overfitting. Training proceeds under an early stopping criterion, with patience of $45$ epochs. The training converges in an average of 150 epochs, running in approximately 10 minutes in a regular CPU environment. The code used in this work can be found on our GitHub\footnote{https://github.com/karlaTame/Neutrino\_SBI/} page.

\section{Results}

\begin{figure}[!h]
 \begin{subfigure}{0.43\textwidth}
     \includegraphics[width=\textwidth]{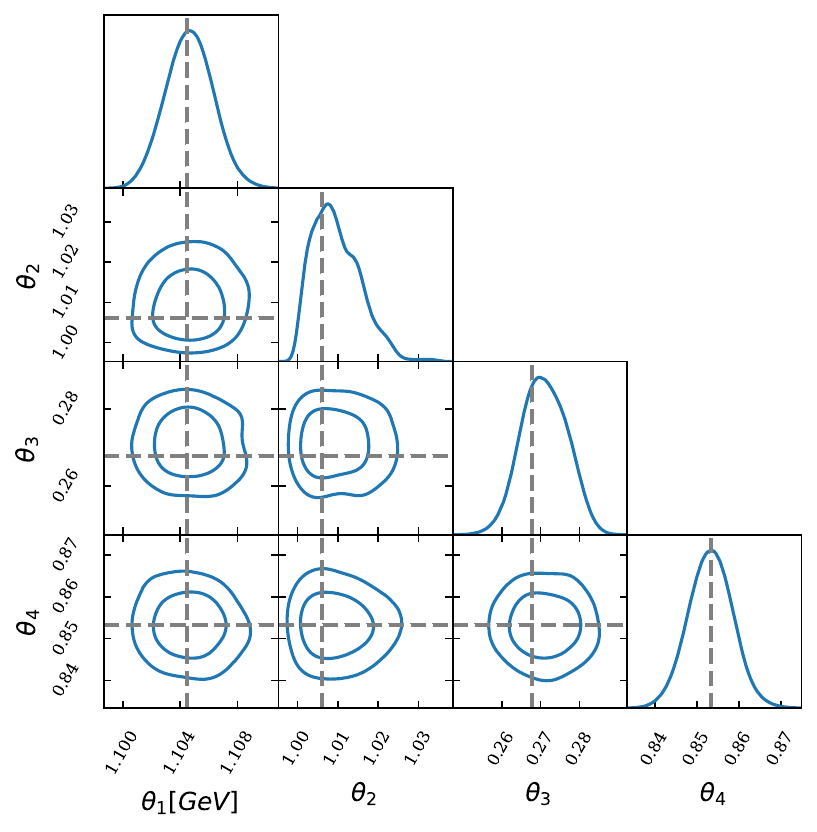}
 \end{subfigure}
 \hfill
 \begin{subfigure}{0.43\textwidth}
     \includegraphics[width=\textwidth]{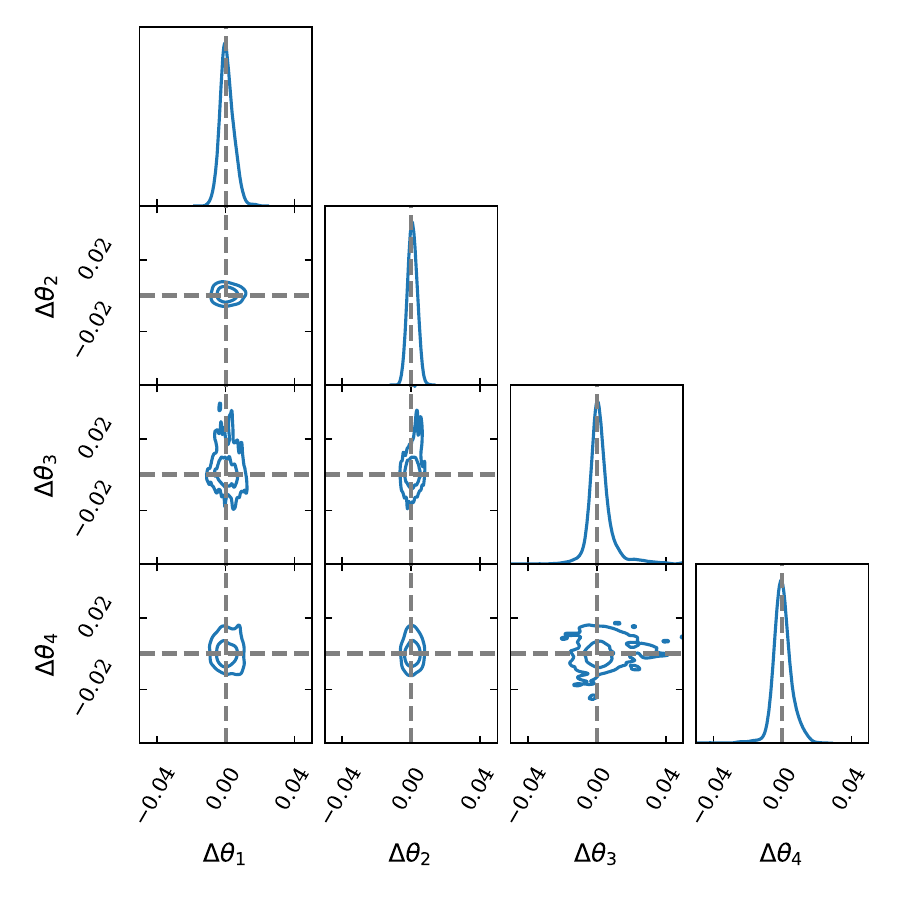}
 \end{subfigure}

 \medskip
 \begin{subfigure}{0.45\textwidth}
     \includegraphics[width=\textwidth]{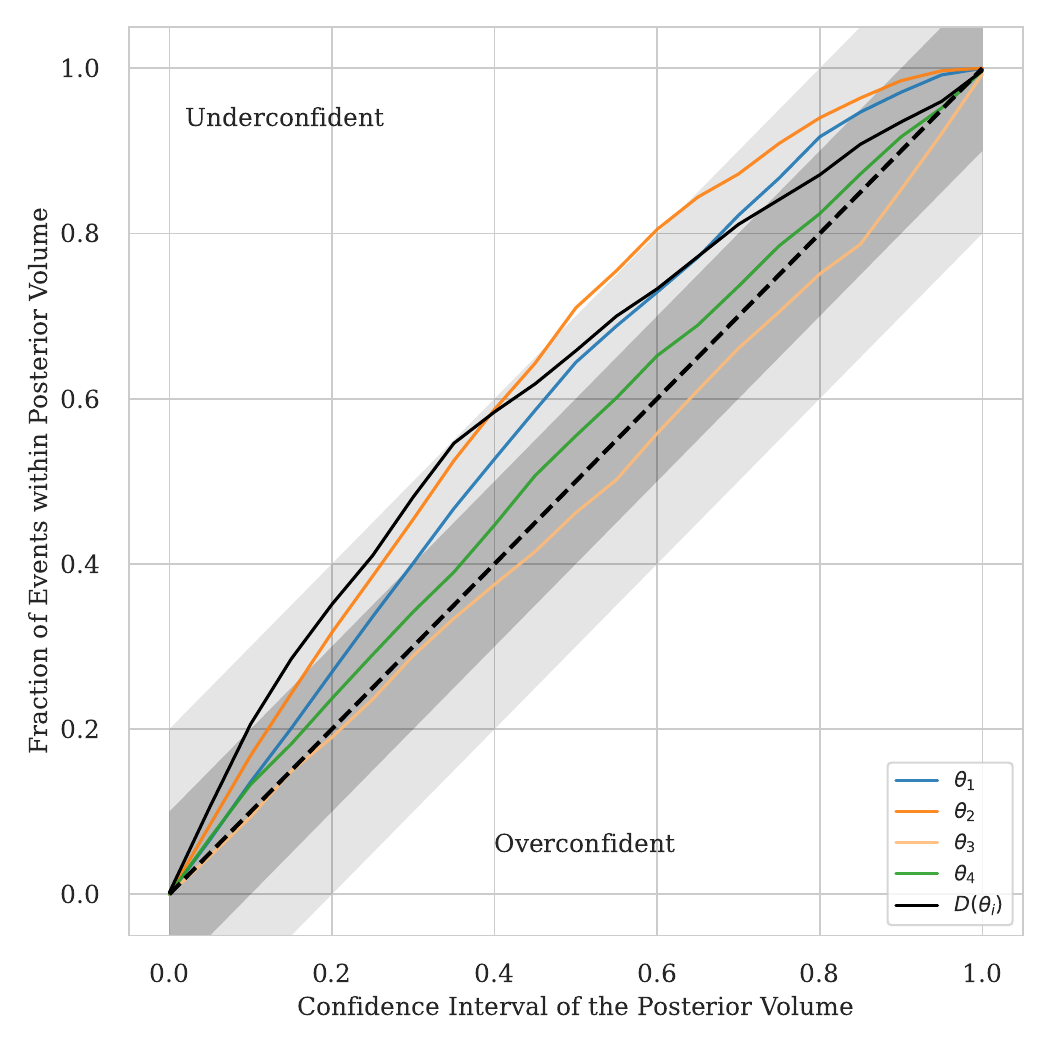}
 \end{subfigure}
 \hfill
 \begin{subfigure}{0.58\textwidth}
     \includegraphics[width=\textwidth]{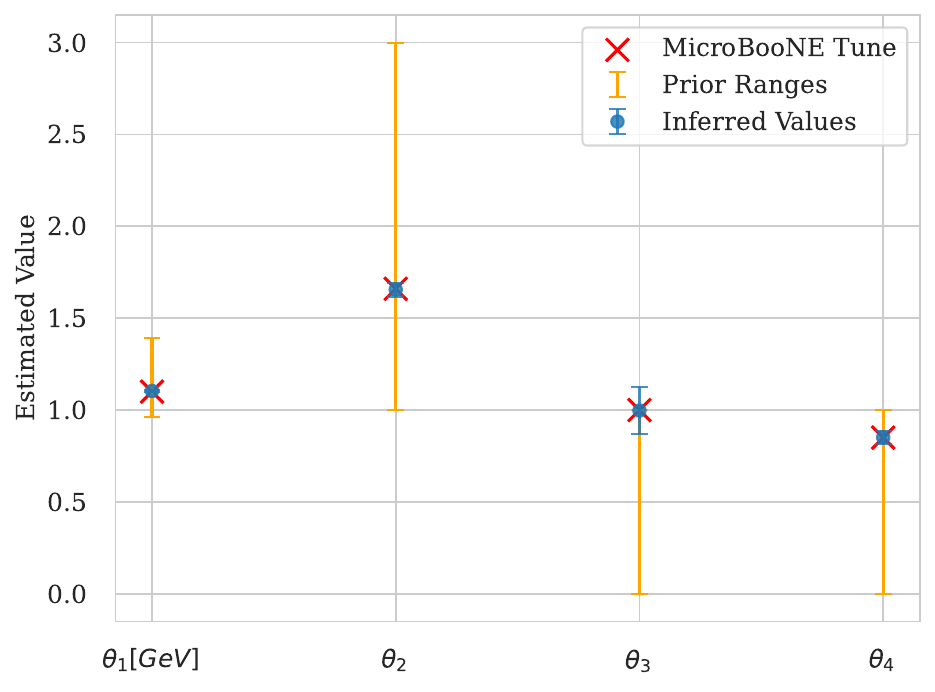}
 \end{subfigure}

 \caption{ Upper left panel: Inferred posterior distributions of four parameters for a single event in the test data set. The gray dashed lines indicate the true values. Upper right panel: Residuals of four parameters for 1000 test events. The gray dashed lines indicate the true value. Lower left panel: Posterior coverage of the $\theta_i$ parameters for $1000$ test events. The diagonal black-dashed line indicates perfect uncertainty calibration. The gray regions indicate thresholds of 10\% (dark gray) and 20\% (light gray) uncertainty miscalibration. Lower right panel: The red points represent the MicroBooNE fit parameters reported in Ref.~\cite{microboonegenietune} and used to generate the test histogram $x_i$. In blue, we show the four parameters $\theta_i$ along with their corresponding $1\sigma$ error bars inferred by our network with $x_i$ as input. In orange, we show the prior ranges used to train the SBI. We observe an excellent match between the inferred and true parameter values.}
 \label{fig1}

\end{figure}

The performance of the NPE for an individual randomly chosen test event is illustrated in Figure~\ref{fig1} (top left figure). In here, the diagonal panels show the one-dimensional marginal posterior distributions for each parameter, while the off-diagonal panels display the corresponding two-dimensional joint posteriors. The contours represent the $68\%$ and $95\%$ confidence intervals, enclosing the highest posterior density intervals under the assumption of a well-behaved posterior distribution. The shape and orientation of the contours provide insight into the correlation structure among the parameters: for example, elongated contours indicate strong correlations, while circular contours suggest independence. In this individual test event, we therefore see little correlation between the four parameters; however, some of these correlations will appear when averaged over the whole test sample (for example, Figure~\ref{fig1} top right panel). The vertical and horizontal dashed lines denote the corresponding true parameter values used to generate the plotted test data point. Overall, the posteriors are centered near the target values within the $68\%$ interval.

To assess the overall performance of the model, in Figure~\ref{fig1} (top right) we show the distribution of the residuals $\Delta\theta_i = \theta_i^{pred} - \theta_i^{true}$ for $i=1,2,3,4$, computed over a sample of $1000$ independent test events. Again, each diagonal panel displays the one-dimensional distribution of residuals for an individual parameter, while the off-diagonal panels illustrate the joint distributions between pairs of residuals. The dashed lines indicate zero bias. We observe that all residuals are centered around zero with narrow widths, indicating that the model yields unbiased estimates with low variance. The contours in the off-diagonal panels reveal the correlation structure among residuals: mild correlations are visible for all parameter pairs, but no significant systematic deviations are observed.

In Figure~\ref{fig1} (bottom left), we present the posterior coverage probability for each parameter $\theta_i$, evaluated again over a sample of 1000 test events. The solid colored lines represent the empirical coverage for each parameter, indicating the fraction of events where the true value lies within a given posterior confidence interval. The black solid curve represents the combined performance of the model on all parameters, for which we utilize a distance metric $D(\theta_i)$ from~\cite{Poh2025}. $D(\theta_i)$ combines parameter values from posterior samples into a single objective function that takes into account the covariance between
different $\theta_i$ parameters. The black dashed diagonal corresponds to perfect uncertainty calibration, where the predicted confidence intervals exactly match the empirical coverage. The shaded bands around the diagonal (dark and light gray) denote 10\% and 20\% tolerance regions, respectively, for quantifying miscalibration. Two of our inferred parameters are within the 10\%, and two within 20\% tolerance band. Only $\theta_3$ shows slight overconfidence, while for the rest of the parameters, the model is slightly underconfident.

Finally, we performed a test to verify the ability of the model to identify the
correct parameter values when presented with a histogram generated with the
MicroBooNE tune parameters. In Figure~\ref{fig1} (bottom right), we show that the true values are almost identically reproduced. This is a key validation
test towards the application of the model to actual experimental data sets,
including the same T2K measurement used to derive the MicroBooNE tune.

\section{Summary and Conclusion}

To overcome present deficiencies in models of neutrino scattering, experimental collaborations often tune simulation parameters to relevant interaction data.
Increasingly arduous requirements for these tuning exercises may be addressed
by improving the sophistication of the numerical methods applied to them. Here we demonstrate, for the first time, the successful application
of advanced AI-based techniques---specifically, simulation-based inference with neural posterior estimation---to efficiently determine the neutrino interaction model
parameter values from mock data. Using outputs of the GENIE and NUISANCE codes, we trained an
SBI algorithm to infer the correct parameter values when presented with a
physics model prediction corresponding to the ``MicroBooNE Tune''
defined in Ref.~\cite{microboonegenietune}. Given that an algorithm trained in this way could be immediately used
for parameter inference using actual measurements as input, this successful validation of our approach lays a strong foundation
for future neutrino interaction model tuning using SBI techniques.

Future work will apply our SBI algorithm to the T2K measurement studied by
MicroBooNE, including a physical interpretation of the inferred parameter
values, a full treatment of correlated uncertainties on the inputs and outputs,
and an evaluation of the consistency and relative performance of our technique
versus the original MicroBooNE likelihood fit. We expect the SBI tuning
approach to be easily generalizable to similar problems in neutrino interaction
modeling, with the potential to significantly improve the efficiency of future
efforts in both time and computing resources.


\begin{ack}
This work was produced by FermiForward Discovery Group, LLC under Contract No. 89243024CSC000002 with the U.S. Department of Energy, Office of Science, Office of High Energy Physics. Publisher acknowledges the U.S. Government license to provide public access under the (\href{http://energy.gov/downloads/doe-public-access-plan}{DOE Public Access Plan}). The work of K.T. is supported by DOE Grant KA2401045.

\end{ack}

\bibliography{sbi}

\begin{thebibliography}{10}

\bibitem{T2K:2016jor}
Ko~Abe et~al.
\newblock {Measurement of double-differential muon neutrino charged-current interactions on C$_8$H$_8$ without pions in the final state using the T2K off-axis beam}.
\newblock {\em Phys. Rev. D}, 93(11):112012, 2016.

\bibitem{microboonegenietune}
P.~Abratenko et~al.
\newblock New {$\mathrm{CC}0\ensuremath{\pi}$ GENIE} model tune for {MicroBooNE}.
\newblock {\em Phys. Rev. D}, 105:072001, Apr 2022.

\bibitem{NOvA:2020rbg}
M.~A. Acero et~al.
\newblock {Adjusting neutrino interaction models and evaluating uncertainties using NOvA near detector data}.
\newblock {\em Eur. Phys. J. C}, 80(12):1119, 2020.

\bibitem{genie2021}
Luis Alvarez-Ruso et~al.
\newblock {Recent highlights from GENIE v3}.
\newblock {\em Eur. Phys. J. ST}, 230(24):4449--4467, 2021.

\bibitem{Anau2023}
Noemi {Anau Montel}, Adam {Coogan}, Camila {Correa}, Konstantin {Karchev}, and Christoph {Weniger}.
\newblock {Estimating the warm dark matter mass from strong lensing images with truncated marginal neural ratio estimation}.
\newblock {\em \mnras}, 518(2):2746--2760, January 2023.

\bibitem{genie2010}
C.~Andreopoulos et~al.
\newblock {The GENIE neutrino Monte Carlo generator}.
\newblock {\em Nucl. Instrum. Methods Phys. Res. A}, 614(1):87--104, 2010.

\bibitem{BM2024}
Ricardo {Barru{\'e}}, Patricia~Conde {Mu{\'\i}{\~n}o}, Valerio {Dao}, and Rui {Santos}.
\newblock {Simulation-based inference in the search for CP violation in leptonic WH production}.
\newblock {\em Journal of High Energy Physics}, 2024(4):14, April 2024.

\bibitem{BB2021}
Sebastian {Bieringer}, Anja {Butter}, Theo {Heimel}, Stefan {H{\"o}che}, Ullrich {K{\"o}the}, Tilman {Plehn}, and Stefan~T. {Radev}.
\newblock {Measuring QCD Splittings with Invertible Networks}.
\newblock {\em SciPost Physics}, 10(6):126, June 2021.

\bibitem{BoeltsDeistler_sbi_2025}
Jan Boelts, Michael Deistler, Manuel Gloeckler, Álvaro Tejero-Cantero, Jan-Matthis Lueckmann, Guy Moss, Peter Steinbach, Thomas Moreau, Fabio Muratore, Julia Linhart, Conor Durkan, Julius Vetter, Benjamin~Kurt Miller, Maternus Herold, Abolfazl Ziaeemehr, Matthijs Pals, Theo Gruner, Sebastian Bischoff, Nastya Krouglova, Richard Gao, Janne~K. Lappalainen, Bálint Mucsányi, Felix Pei, Auguste Schulz, Zinovia Stefanidi, Pedro Rodrigues, Cornelius Schröder, Faried~Abu Zaid, Jonas Beck, Jaivardhan Kapoor, David~S. Greenberg, Pedro~J. Gonçalves, and Jakob~H. Macke.
\newblock sbi reloaded: a toolkit for simulation-based inference workflows.
\newblock {\em Journal of Open Source Software}, 10(108):7754, 2025.

\bibitem{BR2021}
Johann {Brehmer}.
\newblock {Simulation-based inference in particle physics}.
\newblock {\em Nature Reviews Physics}, 3(5):305--305, May 2021.

\bibitem{brehmer2018guide}
Johann Brehmer, Kyle Cranmer, Gilles Louppe, and Juan Pavez.
\newblock A guide to constraining effective field theories with machine learning.
\newblock {\em \prd}, 98(5), September 2018.

\bibitem{Brehmer_2019}
Johann Brehmer, Siddharth Mishra-Sharma, Joeri Hermans, Gilles Louppe, and Kyle Cranmer.
\newblock Mining for dark matter substructure: Inferring subhalo population properties from strong lenses with machine learning.
\newblock {\em \apj}, 886(1):49, nov 2019.

\bibitem{Coogan2022}
Adam {Coogan}, Noemi {Anau Montel}, Konstantin {Karchev}, Meiert~W. {Grootes}, Francesco {Nattino}, and Christoph {Weniger}.
\newblock {One never walks alone: the effect of the perturber population on subhalo measurements in strong gravitational lenses}.
\newblock {\em arXiv e-prints}, page arXiv:2209.09918, September 2022.

\bibitem{CB2020}
Kyle {Cranmer}, Johann {Brehmer}, and Gilles {Louppe}.
\newblock {The frontier of simulation-based inference}.
\newblock {\em Proceedings of the National Academy of Science}, 117(48):30055--30062, December 2020.

\bibitem{gavrikov2025simulationbasedinferenceprecisionneutrino}
A.~Gavrikov, A.~Serafini, D.~Dolzhikov, A.~Garfagnini, M.~Gonchar, M.~Grassi, R.~Brugnera, V.~Cerrone, L.~V. D'Auria, R.~M. Guizzetti, L.~Lastrucci, G.~Andronico, V.~Antonelli, A.~Barresi, D.~Basilico, M.~Beretta, A.~Bergnoli, M.~Borghesi, A.~Brigatti, R.~Bruno, A.~Budano, B.~Caccianiga, A.~Cammi, R.~Caruso, D.~Chiesa, C.~Clementi, C.~Coletta, S.~Dusini, A.~Fabbri, G.~Felici, G.~Ferrante, M.~G. Giammarchi, N.~Giudice, N.~Guardone, F.~Houria, C.~Landini, I.~Lippi, L.~Loi, P.~Lombardi, F.~Mantovani, S.~M. Mari, A.~Martini, L.~Miramonti, M.~Montuschi, M.~Nastasi, D.~Orestano, F.~Ortica, A.~Paoloni, L.~Pelicci, E.~Percalli, F.~Petrucci, E.~Previtali, G.~Ranucci, A.~C. Re, B.~Ricci, A.~Romani, C.~Sirignano, M.~Sisti, L.~Stanco, E.~Stanescu Farilla, V.~Strati, M.~D.~C Torri, C.~Tuvè, C.~Venettacci, G.~Verde, and L.~Votano.
\newblock Simulation-based inference for precision neutrino physics through neural monte carlo tuning, 2025.

\bibitem{GF2021}
Francesca {Gerardi}, Stephen~M. {Feeney}, and Justin {Alsing}.
\newblock {Unbiased likelihood-free inference of the Hubble constant from light standard sirens}.
\newblock {\em \prd}, 104(8):083531, October 2021.

\bibitem{Khullar2022}
Gourav {Khullar}, Brian {Nord}, Aleksandra {{\'C}iprijanovi{\'c}}, Jason {Poh}, and Fei {Xu}.
\newblock {DIGS: deep inference of galaxy spectra with neural posterior estimation}.
\newblock {\em Machine Learning: Science and Technology}, 3(4):04LT04, December 2022.

\bibitem{Legin2021}
Ronan {Legin}, Yashar {Hezaveh}, Laurence {Perreault Levasseur}, and Benjamin {Wandelt}.
\newblock {Simulation-Based Inference of Strong Gravitational Lensing Parameters}.
\newblock {\em arXiv e-prints}, page arXiv:2112.05278, December 2021.

\bibitem{GENIE:2024ufm}
Weijun Li et~al.
\newblock {First combined tuning on transverse kinematic imbalance data with and without pion production constraints}.
\newblock {\em Phys. Rev. D}, 110(7):072016, 2024.

\bibitem{MN2024}
Radha {Mastandrea}, Benjamin {Nachman}, and Tilman {Plehn}.
\newblock {Constraining the Higgs potential with neural simulation-based inference for di-Higgs production}.
\newblock {\em \prd}, 110(5):056004, September 2024.

\bibitem{papamakarios2018maskedautoregressiveflowdensity}
George Papamakarios, Theo Pavlakou, and Iain Murray.
\newblock Masked autoregressive flow for density estimation, 2018.

\bibitem{Poh2025}
Jason {Poh}, Ashwin {Samudre}, Aleksandra {{\'C}iprijanovi{\'c}}, Joshua {Frieman}, Gourav {Khullar}, and Brian~D. {Nord}.
\newblock {Deep inference of simulated strong lenses in ground-based surveys}.
\newblock {\em \jcap}, 2025(5):053, May 2025.

\bibitem{Reza2022}
Moonzarin {Reza}, Yuanyuan {Zhang}, Brian {Nord}, Jason {Poh}, Aleksandra {Ciprijanovic}, and Louis {Strigari}.
\newblock {Estimating Cosmological Constraints from Galaxy Cluster Abundance using Simulation-Based Inference}.
\newblock In {\em Machine Learning for Astrophysics}, page~20, July 2022.

\bibitem{Rubin1984}
Donald~B. Rubin.
\newblock Bayesianly justifiable and relevant frequency calculations for the applied statistician.
\newblock {\em The Annals of Statistics}, 12(4):1151--1172, 1984.

\bibitem{nuisance}
P.~Stowell et~al.
\newblock {NUISANCE: a neutrino cross-section generator tuning and comparison framework}.
\newblock {\em JINST}, 12(01):P01016, 2017.

\bibitem{MINERvA:2019kfr}
P.~Stowell et~al.
\newblock {Tuning the GENIE Pion Production Model with MINER$\nu$A Data}.
\newblock {\em Phys. Rev. D}, 100(7):072005, 2019.

\bibitem{GENIE:2021zuu}
J{\'u}lia Tena-Vidal et~al.
\newblock {Neutrino-nucleon cross-section model tuning in GENIE v3}.
\newblock {\em Phys. Rev. D}, 104(7):072009, 2021.

\bibitem{GENIE:2021wox}
J{\'u}lia Tena-Vidal et~al.
\newblock {Hadronization model tuning in genie v3}.
\newblock {\em Phys. Rev. D}, 105(1):012009, 2022.

\bibitem{GENIE:2022qrc}
Julia Tena-Vidal et~al.
\newblock {Neutrino-nucleus CC0$\pi$ cross-section tuning in GENIE v3}.
\newblock {\em Phys. Rev. D}, 106(11):112001, 2022.

\bibitem{WagnerCarena2022}
Sebastian {Wagner-Carena}, Jelle {Aalbers}, Simon {Birrer}, Ethan~O. {Nadler}, Elise {Darragh-Ford}, Philip~J. {Marshall}, and Risa~H. {Wechsler}.
\newblock {From Images to Dark Matter: End-to-end Inference of Substructure from Hundreds of Strong Gravitational Lenses}.
\newblock {\em \apj}, 942(2):75, January 2023.

\end{thebibliography}

\end{document}